\documentclass[12pt]{article}
\usepackage{amssymb,epsf}
\usepackage{cite}
\usepackage{graphicx}
\newcommand{\beq}{\begin{equation}}
\newcommand{\eeq}{\end{equation}}
\newcommand{\bea}{\begin{eqnarray}}
\newcommand{\eea}{\end{eqnarray}}
\newcommand{\bdm}{\begin{displaymath}}
\newcommand{\edm}{\end{displaymath}}

\newcommand{\noi}{\noindent}
\newcommand{\ra}{\rightarrow}
\renewcommand{\theequation}{\thesection.\arabic{equation}}
\def\<{\langle}
\def\>{\rangle}

\def\a{\alpha}

\def\d{\delta}
\def\e{\epsilon}           
  
\def\g{\gamma}

\def\lam{\lambda}
\def\m{\mu}

\def\p{\pi}                
\def\th{\theta}                   
\def\s{\sigma}                                   

\def\O{\Omega}

\def \R {{\mathbb R}}

\def\N{{\mathcal{N}}}

\begin{document}
\baselineskip=15.5pt
\pagestyle{plain}
\setcounter{page}{1}


\begin{flushright}
{\tt}
\end{flushright}

\vskip 2cm

\begin{center}
{\Large \bf Metallic AdS/CFT}
\vskip 1cm

{\bf Andreas Karch and Andy O'Bannon} \\
\vskip 0.5cm
{\it  Department of Physics, University of Washington, \\
Seattle, WA 98195-1560 \\}
{\tt  E-mail: karch@phys.washington.edu, ahob@u.washington.edu} \\
\medskip

\end{center}

\vskip1cm

\begin{center}
{\bf Abstract}
\end{center}
\medskip
We use the AdS/CFT correspondence to compute the conductivity of massive $\N=2$ hypermultiplet fields at finite baryon number density in an $\N=4$ $SU(N_c)$ super-Yang-Mills theory plasma in the large $N_c$, large 't Hooft coupling limit. The finite baryon density provides charge carriers analogous to electrons in a metal. An external electric field then induces a finite current which we determine directly. Our result for the conductivity is good for all values of the mass, external field and density, modulo statements about the yet-incomplete phase diagram. In the appropriate limits it agrees with known results obtained from analyzing small fluctuations around equilibrium. For large mass, where we expect a good quasi-particle description, we compute the drag force on the charge carriers and find that the answer is unchanged from the zero density case. Our method easily generalizes to a wide class of systems of probe branes in various backgrounds.

\newpage

\section{Introduction} \label{intro}

The anti-de Sitter / conformal field theory (AdS/CFT) correspondence equates the low-energy effective theory of string theory, supergravity, on the background $AdS_5 \times S^5$ with $\N=4$ supersymmetric $SU(N_c)$ Yang-Mills (SYM) theory in the limits of large-$N_c$ and large 't Hooft coupling $\lam = g_{YM} N_c^2$ \cite{Maldacena:1997re}. This conjectured correspondence was originally motivated by the study of solutions of coincident D3-branes in string theory. A finite temperature in the field theory is dual to supergravity in an AdS-Schwarzschild background where the SYM theory temperature $T$ is identified with the Hawking temperature of the AdS black hole \cite{Gubser:1996de,Witten:1998zw}.

The $\N=4$ SYM theory contains fields in the adjoint representation
only. Fields in the fundamental representation may be included by
introducing $\N=2$ hypermultiplets. Introducing a small number $N_f
\ll N_c$ of them, the theory will remain approximately conformal to
leading order in $N_c$ since the beta function goes as $N_f / N_c$.
On the supergravity side this corresponds to introducing D7-branes
and hence open string degrees of freedom \cite{Karch:2002sh}. These
branes are introduced in the probe limit, meaning we have only $N_f
\ll N_c$ of them, and hence the AdS background is left unchanged. In
other words, we neglect the back-reaction of the D7-branes on the
geometry.

Recently, the authors of \cite{Kobayashi:2006sb} initiated a study of this theory at finite baryon number density and in particular began to construct the phase diagram of this theory in the canonical ensemble. The hypermultiplet fields have a global $U(N_f)_V$ vector symmetry and we may identify the $U(1)_B$ subgroup of this as ``baryon number''\cite{Kobayashi:2006sb}. In holography, a global symmetry of the field theory will be dual to some gauge invariance in the gravity theory. In this case a finite baryon density $\langle J^t \rangle$ for $U(1)_B$ current $J^{\m}$ (the exact operator is written in \cite{Kobayashi:2006sb}) is dual on the supergravity side to a nontrivial configuration for the $U(1)$ gauge field on the D7-brane worldvolume.

A simple thought experiment shows how this finite density of hypermultiplet fields behaves similarly to a finite density of electrons in a metal (hence our title). We have a constant $U(1)_B$ charge density $\langle J^t \rangle$ distributed evenly thoughout all of space. If we introduce a non-dynamical, external electric field coupled to baryon number we expect our charge carriers to move in the direction of the applied field. Due to resistance from the $\N=4$ SYM theory plasma the charge carriers will not accelerate forever but will reach a steady state. If the field $E$ points in the $x$ direction, say, we expect a constant, nonzero current $\langle J^x \rangle$ and we can define a conductivity $\s$ by

\beq
\langle J^x \rangle = \s E \nonumber
\eeq

\noi Our goal in this paper is to compute $\s$ using AdS/CFT.

In the field theory the hypermultiplets may be given an $\N=2$ supersymmetry-preserving mass term. This appears in the supergravity theory, for an AdS background without a horizon, as D7-branes that end at some radial position in AdS \cite{Karch:2002sh}. The D7-brane is extended in all the $AdS_5$ directions and wraps an $S^3 \subset S^5$. The position of the $S^3$ on the $S^5$ may be allowed to change with radius and, being a trivial cycle, may collapse to a point. The radial position where this occurs is where the D7-brane ends.

When the background is AdS-Schwarzschild and a horizon is present two topologically distinct classes of D7-brane solutions are possible. The first are analogous to the zero temperature solutions: D7-branes ending outside the horizon. These are called Minkowski solutions (even in Euclidean signature). The second are D7-branes that ``fall into'' the horizon, called black hole solutions. These are D7-branes that fill AdS and intersect the horizon, thus developing a worldvolume horizon. As we change the position where the brane ends we find that the D7-brane will ``jump'' from ending outside the horizon to falling into it. In other words, the D7-brane undergoes a topology-changing transition which by now is very well understood \cite{Babington:2003vm,Kirsch:2004km,Ghoroku:2005tf,Albash:2006ew,Karch:2006bv}. Roughly speaking, the Minkowski solutions are dual to large mass in the boundary theory while black hole solutions are dual to small mass where ``large'' and ``small'' here are relative to the temperature. The topology-changing transition in the bulk appears in the boundary theory as a first-order phase transition that occurs as the hypermultiplet mass is dialed down.

To compute the conductivity we will make great use of two
discoveries of \cite{Kobayashi:2006sb}. The first is that when the
D7-brane worldvolume gauge field corresponding to a finite charge
density is turned on only black hole embeddings are physically
allowed. This means that somehow the black hole solutions alone
``know'' holographically about the entire range of hypermultiplet
masses.\footnote{The first-order transition mentioned above persists
at small density but the line of transitions eventually ends in a
critical point \cite{Kobayashi:2006sb}. More on this below.} How can
black hole solutions encode large mass in the boundary theory? This
was the second important discovery of \cite{Kobayashi:2006sb}. For
large hypermultiplet mass in the boundary theory the D7-brane nearly
resembles a Minkowski solution: it \textit{almost} ends far from the
horizon but develops a ``spike'' that extends all the way down to
the horizon. Instead of collapsing to a point the $S^3$ inside the
$S^5$ stays at a small but constant volume along the spike. In fact
this spike has an action identical to a bundle of strings
\cite{Kobayashi:2006sb}. This makes intuitive sense since a single
quark in the field theory is represented in the supergravity theory
as a single string and a finite density of quarks should appear as
very many strings. What is perhaps surprising is that the D7-brane
alone, with no strings introduced ``by hand,'' realizes such strings
via a spike. In any case, for our analysis of the boundary theory at
finite density we need only consider black hole solutions in the
supergravity theory.

The fact that the induced metric for a black hole solution exhibits a horizon dramatically alters the properties of the brane in the presence of an electric field. A general property of D-branes is that they exhibit an instability
for sufficiently large worldvolume electric field $E$. For a string with both ends on the D-brane a worldvolume $E$-field
will pull the endpoints of the string apart. At a critical value $E_{crit}$ this force grows large enough to overcome the string tension and the string is torn apart, hence the instability. For a black hole embedding any electric field will trigger this instability since $E_{crit}$ goes to zero at a horizon where the vanishing of the time component of the metric makes the string basically tensionless. At finite density we only have black hole embeddings so at finite density this instability is always present.

In the field theory this generic instability is simply due to the fact that at finite charge density any electric field will start to accelerate the charge carriers. In contrast, at zero density the instability is due to pair creation
and only sets in above a nonzero $E_{crit}$ set by the hypermultiplet mass. This nicely illustrates why the Minkowski
embeddings with their nonzero $E_{crit}$ are inappropriate for the boundary theory at finite density. The field theory picture suggests that the endpoint of the instability is a steady state solution with a constant current where the acceleration due to the external electric field is balanced against the drag force experienced by the charge carriers.  It is this steady state solution we find from the bulk point of view in this paper. Consistent with this picture, we will see that the density of charge carriers contributing to the conductivity has two components: those introduced explicitly in $\langle J^t \rangle$ and  those coming from pair creation in the plasma. We will also find that this pair creation is suppressed as the hypermultiplet mass increases, as expected.

So we need to construct D7-brane embeddings corresponding to a field theory plasma with finite baryon number density $\langle J^t \rangle$, constant electric field $E$ in the $x$ direction and a time-independent current $\langle J^x \rangle$. For a finite density $\langle J^t \rangle$ in the boundary theory we need in the supergravity theory a D7-brane worldvolume gauge field with a nontrivial time component $A_t(z)$ for radial coordinate $z$. To accommodate the electric field and the current we further need $A_x(z,t) = - E t + h(z)$ so that we have a constant electric field $F^{tx} = E$ and we require nontrivial $z$ dependence so that we have a nonzero $\langle J^x \rangle$. The behavior of the embedding deep inside AdS-Schwarzschild (near the horizon, roughly speaking) then uniquely fixes $\langle J^x \rangle$ for a given $E$ which allows us to extract the conductivity.

Many transport properties of the $\N=4$ SYM theory alone have been determined using AdS/CFT (see the review \cite{Son:2007vk} and references therein). Of direct interest to us will be the computation of \cite{Caron-Huot:2006te}. By weakly gauging a $U(1)$ subgroup of the R-symmetry the authors of \cite{Caron-Huot:2006te} computed the electrical conductivity of $\N=4$ SYM theory coupled to this $U(1)$. Their result agrees with ours in the appropriate limit as we show in section \ref{conduct}.

The drag force on a single heavy quark moving through the $\N=4$ SYM theory plasma was computed in \cite{Herzog:2006gh,Gubser:2006bz} from stationary string solutions in the AdS-Schwarzschild geometry. We will find that in the same regime of large mass (compared to $\sqrt{\lam} T$) where a good quasi-particle description should be valid we can compute the product $\m M$ where $\m$ is the drag coefficient and $M$ is the kinetic mass (distinct from the Lagrangian mass at finite temperature and density) and find agreement with \cite{Herzog:2006gh,Gubser:2006bz} but only if, as discovered there, we use a \textit{relativistic} relation between velocity and momentum. In particular our result is independent of the density. The conductivity in the large-mass, small external field regime has precisely the form as that in the Drude model of metals.

Everything we do comes with a caveat: the phase diagram of this theory in the full parameter space of $T$, $\langle J^t \rangle$ and  $E$ (in units of the mass) is not complete. We know from \cite{Kobayashi:2006sb} for example that at $E=0$ in the plane of $\langle J^t \rangle$ versus $T$ a region of instabilities does exist so we know the solutions we use are not the true ground state of the system in that region. As found in \cite{Kobayashi:2006sb} the first order transition that occurs at zero density persists at small density but ends in a critical point. The line of first-order transitions is the boundary of the instability region. An assumption throughout this paper of course is that we work at values of $\langle J^t \rangle$ and $T$ outside the instability region. Additionally, as mentioned in \cite{Kobayashi:2006sb}, for sufficiently large density the system could undergo Bose-Einstein condensation, that is, the $U(1)_B$ could be spontaneously broken.\footnote{This could happen because the hypermultiplet contains scalars with Yukawa coupling to fermions and quartic self-coupling and for whom the chemical potential acts as a negative mass squared, allowing a textbook spontaneous symmetry breaking to occur for large enough chemical potential. Using AdS/CFT to determine where this occurs will likely be difficult since the only gauge-invariant observables that could act as order parameters are baryonic operators and dynamical baryons are a difficult problem in AdS/CFT even at zero temperature and density \cite{Kruczenski:2003be}.} If indeed this occurs then our solutions would no longer be the ground state of the system in that phase.

At finite $E$ the phase diagram is currently unknown. We can say little about large electric fields which may trigger new transitions and produce new phases. Our results should be safe in the small-$E$ region, however, where simply by continuity of the first order transitions seen at $E=0$ we expect the phase diagram to be unchanged. More generally, our results are valid whenever the bulk is governed by a D7-brane black hole embedding.

Our method for computing the conductivity actually requires very few ingredients. We need only a valid action for the probe brane, which will be the Dirac-Born-Infeld (DBI) action, and a probe brane worldvolume with a horizon. We may thus generalize our method to a variety of systems of Dq-brane probes in backgrounds of Dp-branes and write down a general formula for the conductivity. We compute one example explicitly, a probe D5-brane in a D3-brane background \cite{Karch:2000gx,DeWolfe:2001pq}, for which the fundamental-representation fields are confined to a (2+1)-dimensional defect.

This paper is organized as follows. In section \ref{probe} we write down the DBI action for our system and solve for the gauge fields. In section \ref{conduct} we compute the conductivity $\s$. In section \ref{drag} we compute $\m M$ in the large-mass limit. In section \ref{dpdq} we generlize our results to Dp/Dq systems.  We conclude with some discussion in section \ref{conclusion}. In the Appendix we compute $\langle J^t \rangle$ and $\langle J^x \rangle$ using holographic renormalization.

\section{The Probe Brane Solution} \label{probe}
\setcounter{equation}{0}

Our $AdS_5$ metric is, in Lorentzian signature and in units where the radius of $AdS$ is one,

\beq
ds^2 = \frac{dz^2}{z^2} - \frac{1}{z^2} \frac{(1 - z^4 / z_H^4)^2}{1+z^4 / z_H^4} dt^2 + \frac{1}{z^2} (1+z^4 / z_H^4) d\vec{x}^2
\eeq

\noi where $z_H^{-1} = \frac{\p}{\sqrt{2}} T$. The boundary is at $z = 0$ and the black hole horizon is at $z = z_H$. Here $d\vec{x}^2$ is the metric of three-dimensional Euclidean space. We will denote the metric components in these directions as $g_{xx}$. Our $S^5$ metric is

\beq
d\Omega_{5}^2 = d\theta^2 + \sin^{2}\theta d\psi^2 + \cos^{2}\theta d\Omega_{3}^2.
\eeq

\noi where $d\Omega_{3}^2$ is the standard metric for an $S^3$ and $\th$ runs from zero to $\p/2$.

As explained in the introduction, we introduce a number $N_f$ of D7-branes filling $AdS_5$ and wrapping the $S^3 \subset S^5$. The DBI action is

\beq
S_{D7} = - N_f T_{D7} \int d^8 \xi \sqrt{- det \left( g_{ab} + (2 \p \a') F_{ab} \right )}
\eeq

\noi plus a Wess-Zumino term that will be zero in what we do. Here $T_{D7}$ is the D7-brane tension, $\xi$ are its worldvolume coordinates, $g_{ab}$ is the induced worldvolume metric and $F_{ab}$ is the worldvolume $U(1)$ field strength (here $a,b$ are worldvolume indices). Our convention is that a string endpoint couples to the worldvolume gauge field with coupling $+1$.

We want an embedding function $\th(z)$ describing the position of the $S^3$ on the $S^5$ as well as worldvolume gauge fields $A_t(z)$ and $A_x(z,t)$. The DBI action becomes

\beq
S_{D7} = - \N \int dz dt \cos^3\th g_{xx} \sqrt{|g_{tt}| g_{xx} g_{zz} - (2 \p \a')^2 \left(  g_{xx} A_t'(z)^2 + g_{zz} \dot{A}_x(z,t)^2 - |g_{tt}| A_x'(z,t)^2 \right )}
\label{original_dbi}
\eeq

\noi Here $g_{zz} = 1/z^2 + \th'(z)^2$. We have divided both sides of this equation by the volume of $\R^3$ so this is an action density. We have also performed the integration over the $S^3$ directions which produces a factor of $2\p^2$. We have included this in the prefactor $\N$, which may be written in terms of boundary theory quantities as

\beq
\N = N_f T_{D7} (2\p^2) = \frac{\lam}{(2 \pi)^4} N_f N_c.
\eeq

\noi where we have used, in our units, $\a'^{-2} = 4 \p g_s N_c = g_{YM}^2 N_c = \lam$. Also important will be the $\lam$-independent quantity

\beq
\N (2 \p \a')^2 = \frac{N_f N_c}{(2 \p)^2}.
\label{gym}
\eeq

\noi If we were to expand the DBI action to quadratic order in the field strength and compare the result to the standard form of the Yang-Mills action $\frac{1}{4 g^2} \int F^2$ then we would identify $1/g^2 = \N (2\p\a')^2$.

We want $A_x(z,t) = -Et + h(z)$. We will therefore have two conserved charges since the action will only depend upon $z$-derivatives of $A_t(z)$ and $h(z)$. The conserved charge associated with $A_t$ is

\beq
\cos^3\th g_{xx} \frac{-\N (2 \p \a')^2 g_{xx} A_t'(z)}{\sqrt{|g_{tt}| g_{xx} g_{zz} - (2 \p \a')^2 \left( g_{xx} A_t'(z)^2 + g_{zz} \dot{A}_x(z,t)^2 - |g_{tt}| A_x'(z,t)^2 \right )}} \equiv D
\eeq

\noi The second charge, associated with $A_x(z,t)$, is

\beq
\cos^3\th g_{xx} \frac{\N (2 \p \a')^2 |g_{tt}| h'(z)}{\sqrt{|g_{tt}| g_{xx} g_{zz} - (2 \p \a')^2 \left( g_{xx} A_t'(z)^2 + g_{zz} \dot{A}_x(z,t)^2 - |g_{tt}| A_x'(z,t)^2 \right )}} \equiv B
\eeq

\noi We can immediately see that $D |g_{tt}| h'(z) = - B g_{xx} A_t'(z)$. Some algebra lets us eliminate $A_t(z)$ and $h(z)$ in favor of $D$, $B$ and $E$. We thus have solutions for the gauge fields

\beq
\label{at}
g_{xx} A_t'(z)^2 = \frac{1}{(2 \p \a')^2} |g_{tt}| D^2 \frac{g_{zz} (|g_{tt}| g_{xx} - (2 \p \a')^2 E^2)}{\N^2 (2 \p \a')^2 |g_{tt}| g_{xx}^3 \cos^6\th + |g_{tt}| D^2 - g_{xx} B^2}
\eeq

\beq
\label{ax}
|g_{tt}| h'(z)^2 = \frac{1}{(2 \p \a')^2} g_{xx} B^2 \frac{g_{zz} (|g_{tt}| g_{xx} - (2 \p \a')^2 E^2)}{\N^2 (2 \p \a')^2 |g_{tt}| g_{xx}^3 \cos^6\th + |g_{tt}| D^2 - g_{xx} B^2}
\eeq

\noi At the horizon, where $|g_{tt}| \ra 0$, we see that the worldvolume magnetic field (a gauge-invariant observable) $F_{zx} = h'(z) \sim |g_{tt}|^{-1/2}$ blows up. This is not a problem, however. The quantity that appears in the action is $|g_{tt}| h'(z)^2$ which goes to $+g_{zz} E^2$ and thus precisely cancels the $- g_{zz} E^2$ term in the DBI action. The other terms in the action vanish individually at the horizon so the action remains finite.

What are the boundary conditions on our gauge fields? Near the $z=0$ boundary the gauge fields asymptotically approach

\beq
A_t(z) = \m - \frac{1}{2} \frac{D}{\N (2\p \a')^2} z^2 + O(z^4)
\label{atasymp}
\eeq

\beq
h(z) = b + \frac{1}{2} \frac{B}{\N (2 \p \a')^2} z^2 + O(z^4)
\label{axasymp}
\eeq

\noi The leading, non-normalizable terms give the sources for the dual operators. $A_t$ is dual to $J^t$ so we interpret $\mu$ as the chemical potential. As in \cite{Kobayashi:2006sb} we require $A_t(z_H) = 0$ which then fixes $D$ in terms of $\mu$. For $h(z)$ we demand simply that the source term $b$ vanishes. The sub-leading, normalizable terms of the asymptotic expansion should give expectation values of the dual operators. In the Appendix we find

\beq
\langle J^t \rangle = D, \qquad \langle J^x \rangle = B
\eeq

\noi To obtain a conductivity we need to determine $\langle J^x \rangle = B$ for given $E$ and $D$. For this we
need to extract a further condition on the solution at large $z$ (in the infrared). We will return to this infrared boundary condition in the next section.

Having solved for the gauge fields we can write the action in terms of $D$, $B$ and $E$ with one dynamical field $\th(z)$,

\beq
S_{D7} = -\N \int dz dt \cos^6\th g_{xx}^{5/2} |g_{tt}|^{1/2} \sqrt{\frac{g_{zz} (|g_{tt}| g_{xx} - (2 \p \a')^2 E^2)}{|g_{tt}| g_{xx}^3 \cos^6\th + \frac{|g_{tt}| D^2 - g_{xx} B^2 }{ \N^2 (2 \p \a')^2 }}}
\label{dbi}
\eeq

\noi We do not obtain the $\th(z)$ equation of motion from this on-shell action, however.  We should either derive the equation of motion from eq. (\ref{original_dbi}) and then plug in the gauge field solutions or Legendre transform to eliminate the gauge fields at the level of the action and then derive the equation of motion. The Legendre transform is

\bea
\bar{S}_{D7} & = & S_{D7} - \int dz dt \left ( F_{zt} \frac{\d S_{D7}}{\d F_{zt}} + F_{zx} \frac{\d S_{D7}}{\d F_{zx}} \right ) \\ & = & -\N \int dz dt \sqrt{\frac{g_{zz}}{|g_{tt}|g_{xx}}} \sqrt{\left ( |g_{tt}| g_{xx} - (2 \p \a')^2 E^2 \right ) \left ( |g_{tt}| g_{xx}^3 \cos^6\th(z) + \frac{|g_{tt}|D^2 - g_{xx}B^2}{\N^2 (2 \p \a')^2} \right )} \nonumber
\eea

\noi which we can check by noting that indeed $\frac{\d \bar{S}_{D7}}{\d D} = A_t'(z)$ and $\frac{\d \bar{S}_{D7}}{\d B} = h'(z)$ as given in eqs. (\ref{at}) and (\ref{ax}).

$\th(z)$ is dual to the hypermultiplet mass operator. The leading, non-normalizable term of $\th(z)$'s asymptotic form gives the mass $m$ of the hypermultiplet fields and the sub-leading, normalizable term gives the expectation value of the mass operator. This expectation value in terms of $\th(z)$'s asymptotic coefficients is written in the Appendix. We will not need this much detail, however. All we will need in what follows is that the zero mass solution is $\th(z)=0$ so that $\cos \th(z) = 1$ while a large mass solution, for the D7-brane nearly ending at the boundary (with a long spike), has $\th(z) \approx \pi/2$ so $\cos \th(z) \approx 0$. The boundary conditions on $\th(z)$ are that $\th(z_H)$ takes some value between zero and $\p/2$ and $\th'(z_H)=0$ as needed for a static solution \cite{Karch:2006bv}.

\section{The Conductivity} \label{conduct}
\setcounter{equation}{0}

As mentioned in the introduction, for a finite density in the SYM theory we only need to consider black hole embeddings in the supergravity theory. This means the $z$-integration in the action $S_{D7}$ of eq. (\ref{dbi}) goes from the $z=0$ boundary to $z=z_H$. Near the horizon, where $|g_{tt}| \ra 0$, both the numerator and denominator under the square root in eq. (\ref{dbi}) are negative. At the boundary both numerator and denominator are positive. The only way for $S_{D7}$ to remain real all the way from $z=z_H$ to $z=0$, then, is if both numerator and denominator change sign at the same special value\footnote{Not surprisingly this is the same argument used in the zero density case of \cite{Herzog:2006gh,Gubser:2006bz} for a single string. In fact in our case we don't even need the effective action: we can make the same argument from the gauge field solutions eqs. (\ref{at}) and (\ref{ax}) where the left-hand sides are manifestly positive for all $z$.} $z=z_*$ defined by the equations

\beq
|g_{tt}|g_{xx} - (2\p\a')^2 E^2 = 0
\label{one}
\eeq

\beq
|g_{tt}| g_{xx}^3 \cos^6\th(z_*) + \frac{|g_{tt}|D^2 - g_{xx}B^2}{\N^2 (2 \p \a')^2} = 0
\label{two}
\eeq

\noi where all metric components are evaluated at $z_*$. These two equations allow us to solve for $z_*$ and in addition
impose one further constraint on the integration constants $B$ and $D$ and hence will allow us to solve for $B$ in terms of $D$ and $E$. This is the infrared boundary condition we have been looking for. We proceed by first solving eq. (\ref{one}) for $z_*$ as a function of $E$:

\beq
z_*^2 = \left ( \sqrt{e^2 + 1} - e \right ) z_{H}^2
\eeq

\noi where we have introduced the dimensionless quantity

\beq
e = \frac{E}{\frac{\p}{2} \sqrt{\lam} T^2}
\eeq

\noi and the signs are chosen to guarantee that $z_*$ is a real number between zero and $z_H$. We will also need $g_{xx}(z_*)$ written in terms of field theory quantities,

\beq
g_{xx}(z_*) = \frac{1}{z_*^2} (1 + z_*^4 / z_H^4) = \p^2 T^2 \sqrt{e^2 + 1}.
\eeq

We use eqs. (\ref{one}) and (\ref{two}) to eliminate $B$ in favor of $D$ and $E$ as

\bea
B^2 & = &  \N^2 (2 \p \a')^2 |g_{tt}| g_{xx}^2 \cos^6 \th + \frac{|g_{tt}|}{g_{xx}} D^2 \\ & = &  \frac{N_f^2 N_c^2 T^2}{16 \p^2} E^2 \sqrt{e^2 + 1} \cos^6 \th + \frac{4}{\p^2 \lam T^4} \frac{E^2 D^2}{e^2 + 1} \nonumber
\eea

\noi where in the second line we have converted to field theory quantities. Identifying $B = \< J^x \>$ we extract the conductivity

\beq
\s = \sqrt{ \frac{N_f^2 N_c^2 T^2}{16 \p^2} \sqrt{e^2 + 1} \cos^6 \th(z_*) + \frac{d^2}{e^2 + 1} }
\label{sigma}
\eeq

\noi where we define $d$ similarly to $e$,

\beq
d = \frac{D}{\frac{\p}{2} \sqrt{\lam} T^2} = \frac{\< J^t\>}{\frac{\p}{2} \sqrt{\lam} T^2}
\eeq

\noi but notice that while $e$ was dimensionless $d$ has dimension one.

Eq. (\ref{sigma}) is our main result. It has a simple interpretation. Two types of charge carriers contribute to the conductivity. One type comes from the charge carriers we have introduced explicitly, represented by the $d^2$ term. The other type comes from charge carriers thermally produced in charge-neutral pairs. The effect of these thermally produced pairs should be Boltzmann suppressed at large mass. Indeed, our conductivity depends upon the mass of the fundamental, microscopic charge carriers via the $\cos \th$ term: taking $m \ra \infty$ sends $\cos \th \ra 0$ which obviously reduces $\s$ while taking $m \ra 0$ sends $\cos \th \ra 1$ which obviously enhances $\s$. This is consistent with the fact that lighter particle/anti-particle pairs can be more easily excited thermally.

When the mass is large the $\cos \th$ term can be neglected and the conductivity is proportional to the density of charge carriers $\< J^t \>$ as expected in a quasi-particle interpretation. In this regime we can compare to the work of \cite{Herzog:2006gh}. We will return to this in the next section.

On the other hand, in the limit of zero mass, zero density and zero external field we can compare our ``macroscopic'' result for the conductivity with the ``microscopic'' answer of \cite{Caron-Huot:2006te} where the conductivity was determined from the study of small fluctuations in equilibrium via a Kubo formula. In \cite{Caron-Huot:2006te} $\s$ was the conductivity of the pure $\N = 4$ SYM theory plasma and a gauged $U(1)$ R-symmetry played the role of electromagnetism. The answer of \cite{Caron-Huot:2006te} may be written as $\s = \p T / g^2$ where $g^2$ is the coupling associated with the bulk Yang-Mills Lagrangian, defined in \cite{Policastro:2002se} as $g^2 = 16 \p^2 / N_c^2$. In our case we find in this limit

\beq
\s = \frac{N_f N_c T}{4 \p}.
\eeq

\noi For us the Yang-Mills Lagrangian comes from expanding the DBI action to quadratic order in the gauge field strength. As explained below eq. (\ref{gym}) this is $1/g^2 = \N (2\p\a')^2 = N_f N_c / 4 \p^2$ so indeed our answer\footnote{Notice also that our coupling between the current and the external vector potential is one so their $e^2$ factor does not appear in our answer.} is $\p T / g^2$.

\section{The Drag Force} \label{drag}
\setcounter{equation}{0}

Now we return to nonzero density and external field. We will take the limit where the mass is much larger than $\sqrt{\lam}T$ \cite{Herzog:2006gh}, for which $\cos \th \approx 0$. In this case we expect a good quasi-particle description. We should thus be able to compare our result to that of \cite{Herzog:2006gh}, who wrote the equation of motion for the quasi-particles

\beq
\frac{dp}{dt} =  - \m p + f.
\eeq

\noi The external force in our conventions is $f = E$. Here $\m$ is the friction (or drag) coefficient, not the chemical potential. We look at the equilibrium case $\frac{dp}{dt} = 0$.  In order to compare to \cite{Herzog:2006gh} we employ a relativistic relation between mass and momentum,

\beq
\label{relativistic}
\m M \frac{v}{\sqrt{1-v^2}} = E
\eeq

\noi for kinetic mass $M$. In \cite{Herzog:2006gh} it was found that the quasi-particle obeys this relativistic relation despite the fact that its rest mass and its kinetic mass are not the same. In order to extract $\m M$ we need to compute $v$ as a function of $E$.

At large mass we expect pair creation to be suppressed so we expect only the charge carriers in $\langle J^t \rangle$ to contribute to $\langle J^x \rangle$. We may thus write $\langle J^x \rangle = \langle J^t \rangle v$ where $v$ is the velocity of the quasi-particles. Equating this with $\< J^x \> = \s E$ we have $v = \s E/\< J^t \>$. At large mass we take $\cos \th \approx 0$ so we can ignore the first term under the square root in eq. (\ref{sigma}) to find

\beq
v = \frac{d}{\sqrt{e^2 + 1}} \frac{E}{\< J^t \>} = \frac{e}{\sqrt{e^2 + 1}}, \qquad \frac{v}{\sqrt{1-v^2}} = e
\eeq

\noi We find simply

\beq
\m M = \frac{E}{e} = \frac{\p}{2} \sqrt{\lam} T^2,
\eeq

\noi This answer agrees with that of \cite{Herzog:2006gh}, who highlighted the mass independence of this result. We have found that at nonzero density the result is also independent of the density.

Having identified $\m M$ in the large-mass limit we can in addition take the small $e$ limit of eq. (\ref{sigma}) and find

\beq
\s = \frac{\langle J^t \rangle}{\m M}
\eeq

\noi which is precisely the form of $\s$ in the Drude theory of metals, $\s = n e / \mu m$ for electrons of charge $e$, density $n$ and mass $m$. Our field theory of very massive charge carriers in a weak field behaves very much like a Drude metal. This comparison also shows that the non-linear effects in $e$ can be explained in terms of the pseudo-relativistic behavior of the quasi-particles displayed in eq. (\ref{relativistic}).

\section{Generalization to Dp/Dq Systems} \label{dpdq}
\setcounter{equation}{0}

A very similar analysis can be applied to $N_f$ probe Dq-branes in the background of $N_c$ Dp-branes for which the holographic duals will be flavored field theories, possibly with the flavor fields confined to a defect\footnote{At zero temperature these systems for general p and q were analyzed in \cite{Arean:2006pk,Myers:2006qr}.}. Our method is applicable to these systems because we required only that the DBI action be a reliable effective action and that the probe brane had a worldvolume horizon (with the associated boundary conditions on the worldvolume gauge fields). We will perform a general analysis and then provide one example, the probe D5-brane in a background of D3-branes.

The coordinates of a black Dp-brane (p$<7$) solution include coordinates parallel to the Dp-branes and spherical coordinates for the directions transverse to the Dp-branes. In this background the induced Dq-brane metric will generically be

\beq
ds_{Dq}^2 = g_{zz} dz^2 + g_{tt} dt^2 + g_{xx} d\vec{x}^2 + g_{SS} d \O_n^2
\eeq

\noi where $z$ is the radial coordinate. We assume the metric depends only on $z$ and parameters like $T$. The Dq-brane wraps some $n$-sphere $S^n$ in the space transverse to the Dp-brane worldvolumes. Here $g_{SS}$ is the metric component on this sphere. The Dq-brane worldvolume then includes some $\R^{d}$ with metric components $g_{xx}$ where $d=q-n-1$ .  We assume the Dq-brane worldvolume has a horizon $z_H$ defined by $g_{tt}(z_H) = 0$. We also hide any embedding information (such as our $\th(z)$ above) in the components of this induced metric. The Dp-brane background will also generally include a nontrivial dilaton $\phi(z)$.

We now introduce $A_t(z)$ and $A_x(z,t) = - E t + h(z)$ with the usual boundary conditions. The Dq-brane action is then

\beq
S_{Dq} = - N_f T_{Dq} V_n \int dz dt e^{-\phi} g_{xx}^{(d-1)/2} g_{SS}^{n/2} \sqrt{g_{zz} g_{xx}|g_{tt}| - (2\p\a')^2 (g_{xx}A_t'^2 + g_{zz}\dot{A}_x^2 - |g_{tt}|A_x'^2)}
\eeq

\noi where $T_{Dq}$ is the Dq-brane tension and $V_n$ is the volume of a unit $S^n$. We have divided both sides by the volume of $\R^{d}$. Comparing to eq. (\ref{original_dbi}) we can see that everything will be identical to what we have already done but with the replacements

\beq
\N \ra \N_q \equiv N_f T_{Dq} V_n
\eeq

\beq
g_{xx} \cos^3\th \ra e^{-\phi} g_{xx}^{(d-1)/2} g_{SS}^{n/2}
\eeq

\noi In particular $\langle J^t \rangle = D$ and $\langle J^x \rangle = B$ are still true (see the Appendix). We may jump to the equations

\beq
|g_{tt}| g_{xx} - (2\p\a')^2 E^2 = 0
\label{onedpdq}
\eeq

\beq
e^{-2 \phi} g_{xx}^{d} g_{SS}^{n} |g_{tt}| + \frac{|g_{tt}| D^2 - g_{xx} B^2}{\N_q^2 (2\p\a')^2} = 0
\label{twodpdq}
\eeq

\noi both evaluated at $z_*$. We again construct $\s = \< J^x \> / E = B / E$,

\beq
\s = \sqrt{ \N_q^2 (2\p\a')^4 e^{-2\phi} g_{xx}^{d-2} g_{SS}^n + (2\p\a')^2 g_{xx}^{-2} D^2 }.
\eeq

\noi This is the general form. To go beyond this requires choosing a specific system.

For this $\s$ the value $d=2$ is clearly special: the $g_{xx}$ factor in the first term drops out. Upon translating to boundary theory quantities this means some non-linearities in $E$ are dropping out.\footnote{This is also the right dimension to have interesting phenomena, such as the quantum Hall effect, govern the conductivity.} This leads us to the example of a probe D5-brane in the same D3-brane background we have considered above. In this case the dual theory is $\N=4$ SYM theory in $3+1$ dimensions coupled to $N_f$ massive fields in the fundamental representation that are confined to a (2+1)-dimensional defect. The D5-brane wraps $AdS_4 \times S^2$ inside $AdS_5 \times S^5$. We have $d=2$, $n=2$ , $\phi(z) = 0$ and the $S^2$ metric component $g_{SS} = \cos^2 \th(z)$ has the same interpretation as for the D7-brane. We find

\beq
\s = \sqrt{\frac{4 N_f^2 N_c^2}{\p^2 \lam} \cos^4\th(z_*) + \frac{d^2}{e^2 + 1}}
\eeq

\noi where $d$ and $e$ are defined as before but now $d$ is dimensionless ($D$ has dimension two in 2+1 dimensions) so $\s$ is dimensionless as it should be. Notice that taking zero density and zero mass gives a constant so in this limit $\< J^x \>$ is purely linear in $E$. We may compare this result to that of \cite{Herzog:2007ij} where $\s$ was the conductivity of $\N=8$ SYM theory in (2+1) dimensions and a $U(1)$ subgroup of the $SO(8)$ R-symmetry played the role of electromagnetism. The result of \cite{Herzog:2007ij} was $\s = 1/ g^2$ where $g$ was the coupling of the $U(1)$ Yang-Mills theory formulated on $AdS_4$. Again we idenfity $g$ in our case by expanding the DBI action of the D5-brane to quadratic order in the field strength to find $1 / g^2 = \N_5 (2 \p \a')^2 = 2 N_f N_c / \p \sqrt{\lam}$ so indeed our answer in this limit is $\s = 1/ g^2$.

\section{Discussion and Conclusion} \label{conclusion}

We have computed the conductivity of a finite density of massive $\N=2$ hypermultiplets in an $\N=4$ SYM theory
plasma. In contrast to earlier work using AdS/CFT to study transport coefficients our approach is ``macroscopic'' in the sense that we directly calculate the response (in our case the current) as a result of a large external perturbation of the plasma (the electric field). This method nicely complements the existing studies which relate transport to small fluctuations in equilibrium via the Kubo formulas.

For massless flavors and zero density our result completely agrees with answers from the small fluctuation analysis. For small but finite masses a similar comparison to a small fluctuation analysis performed for a zero density black hole embedding should be possible but has not been performed yet. At large mass (compared to $\sqrt{\lambda} T$), where the zero density embedding for the flavor brane is a Minkowski embedding, the classical small fluctuation analysis would give $\sigma=0$ since the induced metric on the flavor brane has no horizon and hence no purely outgoing boundary conditions, which are the source of dissipative phenomena in the bulk, could be imposed. Only quantum fluctuations of the flavor
brane could reveal the conductivity in that case. Our answer encompases all of these cases.

We emphasize that we have worked at leading order in $N_f/N_c$ at large $N_c$, or in supergravity language, we have not included the backreaction of the D7-branes. If we were to do so the corresponding solution would no longer be static. The external field $E$ is pumping energy and momentum into the system at a finite rate so the total energy and momentum have
to grow linearly in time. At first it seems puzzling how the gravitational backreaction of the D7-brane should see this. The stress tensor associated with the DBI action only depends on the gauge invariant field strength (not on the vector potential itself) and so is completely static for our solution. One can find a solution to Einstein's equation with this source that is completely static. This solution has to be unphysical due to the boundary conditions imposed
at the horizon. It is well known that close to the horizon at least the linearized Einstein's equations reduce to a standard wave equation. For $e^{i \omega t}$ time dependence one demands as a physical boundary condition that the wave is purely outgoing, that is, nothing comes back into the physical space through the horizon. For zero $\omega$ one has to be more careful: both a constant and a solution linear in time are possible. The correct boundary condition
on the horizon capturing the physics of the field theory must pick the solution linear in time.

For the stationary solution $\langle J^x \rangle = \sigma E$ to be valid the charged quasiparticles need to dissipate their momentum so we can balance the drag force against the external force from the  external field. If the densities of charged ($\N=2$ hypermultiplet) and uncharged ($\N=4$ SYM theory) particles were comparable, momentum conservation would dictate that the drag force that allowed the charge carriers to dissipate their momentum would at the same time accelerate the uncharged parts of the plasma and no stationary stage could be reached. In other words, the hypermultiplet fields would begin to drag the $\N=4$ SYM theory plasma along with them, which is clearly not a stationary solution. In our case however only the fundamental-representation fields carry charge so that the energy density of charge carriers is of order $N_f N_c$ while the energy density of the neutral part of the plasma is of order $N_c^2$. The charge carriers can dissipate an order $N_c$ momentum density at a constant rate without causing any meaningful velocities in the ${\cal N}=4$ SYM theory plasma. Only after time $t \sim N_c$ will the momentum in the neutral plasma have built up to order $N_c^2$ and hence give velocities of order one. At such late times a simple description of the form $\langle J^x \rangle = \sigma E$ will break down. In this way the background ${\cal N}=4$ plasma at large $N_c$ acts like the lattice in solid state physics (again suggesting out title): it can absorb an arbitrary amount of momentum without experiencing significant macroscopic motion. Unlike the lattice, it does so simply by its large density rather than by breaking translational invariance.

For the future one should be able to use our macroscopic approach to get a better understanding of transport properties of various strongly coupled systems via AdS/CFT. Especially promising seems to be the application to the (2+1)-dimensional case where conductivities in the presence of magnetic fields could exhibit quantum Hall behavior.

\section*{Acknowledgements}

We would like to thank C. Herzog, D.T. Son and L. Yaffe for helpful discussions.
The work of A.K. was supported in part by DOE contract \# DE-FG02-96-ER40956.
The work of A. O'B. was supported by the Jack Kent Cooke Foundation.

\section*{Appendix: Holographic Renormalization} \label{holorg}
\setcounter{equation}{0}
\renewcommand{\theequation}{A-\arabic{equation}}

The AdS/CFT dictionary equates the on-shell action $S_{D7}$ with the generating functional of field theory correlation functions. The on-shell action, however, is divergent due to integration over the infinite volume of AdS space, that is, due to integration all the way to the boundary at $z=0$. Holographic renormalization (holo-rg) \cite{Henningson:1998gx,Henningson:1998ey,Balasubramanian:1999re,deHaro:2000xn} regulates the divergence by introducing a cutoff $z=\e$ and then adding counterterms on the $z=\e$ slice to cancel divergences before taking $\e \ra 0$.

We find (from its equation of motion) that $\th(z)$ has an asymptotic expansion

\beq
\th(z) = \th_0 z + \th_2 z^3 + \ldots.
\eeq

\noi Plugging this asymptotic form into the regulated action we find the divergences

\beq
S_{reg} = - \int_{\e}^{z_H} {\cal L} = - \N \int_{\e}^{z_H} dz \left ( z^{-5} - \th_0^2 z^{-3} - \frac{1}{2} (2 \p \a')^2 E^2 z^{-1} + O(z) \right )
\eeq

\noi The first two terms are clearly divergent as $\e \ra 0$. The counterterms we need to cancel these divergences are \cite{Karch:2005ms}

\beq
L_1 = \frac{1}{4} \N \sqrt{-\g}, \qquad L_2 = - \frac{1}{2} \N \sqrt{-\g} \th(\e)^2, \qquad L_f = \N \frac{5}{12} \sqrt{-\g} \th(\e)^4
\eeq

\noi where $\g_{ij}$ is the induced metric on the $z=\e$ slice and $\g$ is its determinant. Notice that $\sqrt{-\g} = \e^{-4} + O(\e^4)$. As explained in \cite{Karch:2005ms} supersymmetry requires the third counterterm, which is finite.

In all of the equations above we have suppressed a factor of $\int dt$. We will continute to do so until we compute $\langle J^x \rangle$ since only $A_x$ has time dependence. The last divergence requires a counterterm

\bea
L_F & = & - \frac{1}{4} \N (2 \p \a')^2 \sqrt{-\g} F^{ij} F_{ij} \log \e \\ & = & - \frac{1}{4} \N (2 \p \a')^2 \sqrt{-\g} \g^{ij} \g^{kl} F_{ik} F_{jl} \log \e \nonumber \\ & = & + \frac{1}{2} \N (2 \p \a')^2 E^2 \log \e + O(\e^4 \log \e) \nonumber
\eea

The generating functional of the boundary theory is then the $\e \ra 0$ limit of $S_{D7} = S_{reg} + \sum_i L_i$. We interpret the leading coefficient $\th_0$ as the source for the dual operator, which we denote as\footnote{The exact operator is written in \cite{Kobayashi:2006sb}.} $\langle \bar{\psi} \psi \rangle$. In other words we equate $\th_0$ with the mass\footnote{Up to a normalization explained in section 3.2 of \cite{Karch:2006bv}.} of the fundamental-representation fields. The expectation value of this operator is \cite{Karch:2005ms}

\beq
\langle \bar{\psi} \psi \rangle = \lim_{\e \ra 0} \frac{1}{\e^3} \frac{1}{\sqrt{-\g}} \frac{\d S_{D7}}{\d \th(\e)} = - \N \left ( - 2 \th_2 + \frac{1}{3} \th_{0}^3 \right ).
\eeq

We need to compute expectation values of the $U(1)$ current components $\langle J^t \rangle$ and $\langle J^x \rangle$. In the formalism of holographic renormalization $\langle J^t \rangle$ is

\beq
\langle J^t \rangle = \lim_{\e \ra 0} \frac{1}{\e^4} \frac{1}{\sqrt{-\g}} \frac{\d S_{D7}}{\d A_t(\e)}
\eeq

\noi Making arguments similar to those of \cite{Kobayashi:2006sb}, we have

\beq
\d S_{D7} = - \int_{\e}^{z_H} dz \frac{\d \cal L}{\d \partial_z A_t} \partial_z \d A_t = - D \int_{\e}^{z_H} dz \partial_z \d A_t  = -D \left ( \d A_t(z_H) - \d A_t(\e) \right ).
\eeq

\noi We enforce the boundary condition $\d A_t(z_H)=0$ at the horizon. We thus find $\frac{\d S_{D7}}{\d A_t(\e)}= D$ and hence $\langle J^t \rangle = D$.

The density $\langle J^x \rangle$ is slightly more subtle because $A_x$ is time dependent. In holo-rg we have

\beq
\langle J^x \rangle = \lim_{\e \ra 0} \frac{1}{\e^4} \frac{1}{\sqrt{-\g}} \frac{\d S_{D7}}{\d A_x(\e)}
\eeq

\noi We now have two terms (reinstating $\int dt$)

\beq
\d S_{D7} = -\int dt dz \left ( \frac{\d \mathcal{L}}{\d \partial_z A_x} \partial_z \d A_x + \frac{\d \mathcal{L}}{\d \partial_t A_x} \partial_t \d A_x \right ).
\eeq

\noi We employ the same argument as before for the first term,

\beq
-\int_{\e}^{z_H} dz \frac{\d \mathcal{L}}{\d \partial_z A_x} \partial_z \d A_x = -B \left ( \d A_x(z_H) - \d A_x(\e) \right ) = B  \d A_x(\e).
\eeq

\noi In the second term $\frac{\d \mathcal{L}}{\d \partial_t A_x}$ is $t$-independent on-shell and hence

\beq
-\int dt dz \frac{\d \mathcal{L}}{\d \partial_t A_x} \partial_t \d A_x = - \int dz \frac{\d \mathcal{L}}{\d \partial_t A_x} \int dt \partial_t \d A_x = 0
\eeq

\noi where we have demanded that the fluctuation be well-behaved (vanishing) at $t = \pm \infty$ so that $\int dt \partial_t \d A_x = \d A_x(+\infty) - \d A_x(-\infty) =  0$. The counterterm $L_F$ gives a vanishing contribution to $\langle J^x \rangle$ for the same reason

\bea
\d L_F & = & - \frac{1}{4} \N (2 \p \a')^2 \sqrt{-\g} \g^{ij}\g^{kl} \int dt \frac{\d }{\d \partial_t A_x} \left( F_{ik} F_{jl} \right) \partial_t \d A_x \log \e \\ & = & + \frac{1}{2} \N (2 \p \a')^2 \dot{A}_x(\e) \log \e \int dt \partial_t \d A_x + O(\e^4 \log \e) \nonumber \\ & = & O(\e^4 \log \e) \nonumber
\eea

\noi We then have $\frac{\d S_{D7}}{\d A_x(\e)}= B$ and hence $\langle J^x \rangle = B$.

In regard to section \ref{dpdq} we can see that these results will be valid for any probe Dq-brane. The identification of $\langle J^t \rangle = D$ depended only on the behavior of $A_t(z)$ in the radial direction, which will be unchanged for the class of systems we considered (probe branes with worldvolume horizons). A similar statement applies for the identification $\langle J^x \rangle = B$. The one subtlety is that additional counterterms may appear for different systems. No such counterterms can change these results, however. Any counterterm on the $z=\e$ slice must be built from gauge- and Lorentz-invariant combinations of the field strength. The only component of the field strength that could contribute would be $F_{tx}$ (and not $F_{zt}$ or $F_{zx}$). The crucial point is that $F_{tx} = -E$ is time-independent so we will always end up with $\int dt \partial_t \d A_x = 0$ as above.

\end{document}